# Highly efficient wideband and polarization-insensitive SMF-ARF coupling strategy with low back-reflection

Yi Su[1†], Xuchen Hua[2†], Bingyan Xue[1], Yucheng Yao[2], Zhiyong Zhao[2], and Ming Tang[1,2*]

[1]School of Future Technology, Huazhong University of Science and Technology, Wuhan 430074, China
[2]School of Optical and Electronic Information, Huazhong University of Science and Technology, Wuhan 430074, China

†These authors contributed equally.
*Corresponding author: tangming@mail.hust.edu.cn

We propose a lensed-fiber based coupling strategy for low-loss interconnection between single-mode fibers and anti-resonant fibers. By optimizing structural and geometric parameters, the design simultaneously achieves high coupling efficiency and suppressed back-reflection. Experimental results demonstrate an insertion loss of 1.2 dB and back-reflection of -36.22 dB at 1550 nm, with excellent spectral stability (below 0.72 dB variation across 1500-1600 nm) and polarization insensitivity (below 0.4 dB polarization dependent loss). The compact structure not only facilitates the fabrication process, but also enables seamless ARF integration into existing optical networks, thereby addressing critical demands for high-capacity data transmission.
DOI: 10.3788/COLXXXXXX.XXXXXX.

## 1. Introduction

The exponential advancement of artificial intelligence (AI), driven by the intensifying computational demands of applications such as artificial general intelligence[1], cloud computing[2], and virtual reality[3], has precipitated unprecedented growth of hyper-scale datacenters[4-6], which urgently necessitates the expansion of optical communication capacity. As the backbone of modern data center infrastructure, optical communication systems now face critical challenges in meeting the escalating requirements for reliability, ultra-low latency (<100 ns), and terabit-scale bandwidth[7]. However, the intrinsic Rayleigh scattering, dispersion, and nonlinearity of silica fibers limit their bandwidth and loss, restricting capacity and efficiency[8]. Recently, anti-resonant fibers (ARFs) have attracted increasing interest due to their ultra-low nonlinearity, and low group velocity dispersion[8], which are excellent candidates for substituting traditional SMFs. Given that the majority of existing optical fiber infrastructure is predominantly constructed with SMFs, achieving low-loss interconnections between ARFs and single mode fibers (SMFs) has become a critical imperative in photonic systems engineering[12].

Current SMF-ARF interconnection techniques face several critical challenges. Direct fusion splicing often causes ARF structural deformation, resulting in excessive insertion loss exceeding 3 dB[13]. Fiber tapering methods, while effective for ARFs with core diameters larger than 50 μm, show limited performance with smaller-core ARFs due to mode-field mismatch[14]. More fundamentally, these approaches fail to adequately address Fresnel reflections at air-glass interfaces. Although angle-cleaved fusion splicing can reduce reflections approximately by 15 dB[15], the angled interface introduces beam deviation greater than 2 degrees, leading to higher-order mode excitation and additional loss. Recent advances using GRIN fiber adapters with anti-reflection coatings have demonstrated improved performance where insertion losses below 1.5 dB while maintaining reflection levels under -40 dB [16,17], but the stringent requirements on GRIN fiber length control within ±10 μm and complex coating processes pose significant manufacturing challenges for practical deployment.

In this paper, an easy-to-manufacture coupling approach for ARFs employing lensed SMFs is proposed and fabricated, with simultaneously optimized coupling efficiency and back-reflection performance via structural design. Systematic numerical simulations investigated the influence of key geometric parameters on optical performance, with experimental validation demonstrating an insertion loss (IL) of 1.2 dB and back-reflection below -36.22 dB. The fabricated structure combines simplicity and compactness with high performance, exhibiting spectral stability within 0.72 dB across the 1500–1600 nm wavelength range and polarization insensitivity with a polarization dependent loss (PDL) below 0.4 dB. These combined advantages enable direct integration of hollow-core ARFs into existing SMF-based infrastructure, providing a practical solution for upgrading optical communication systems without requiring complex retrofitting.

## 2. Principle

The schematic diagram of proposed coupling method is shown in Fig. 1(a). This method involves only a standard SMF and an ARF, both of which are precisely fixed within a heat-processed glass tube. The core and cladding diameters of the SMF are 9 μm and 125 μm, respectively. Additionally, its end is lensed to adapt the mode fields between the SMF and the ARF. The microscopic structural diagram of the ARF is demonstrated in Fig. 1(b), exhibiting that the ARF comprises five circularly symmetrical, non-contact nested anti-resonant tubes. Specifically, the inner tube has a diameter of 15.23 μm while the outer tube has a diameter of 26.95 μm. The core and cladding diameters of the ARF are 87.77 μm and 240 μm.

Coupling loss arises principally from mode-field mismatch between ARF and SMF, accompanied by Fresnel reflections at dielectric interfaces[18]. The predominant factor, mode-field mismatch, originates from fundamental discrepancies in both mode-field size and shape. Specifically, a significant difference exists in the mode field diameters (MFDs), as evidenced by finite element simulations performed in COMSOL Multiphysics, which yields fundamental mode MFDs of 23.2 μm for the ARF and 8.4 μm for the SMF, as illustrated in Fig. 2.

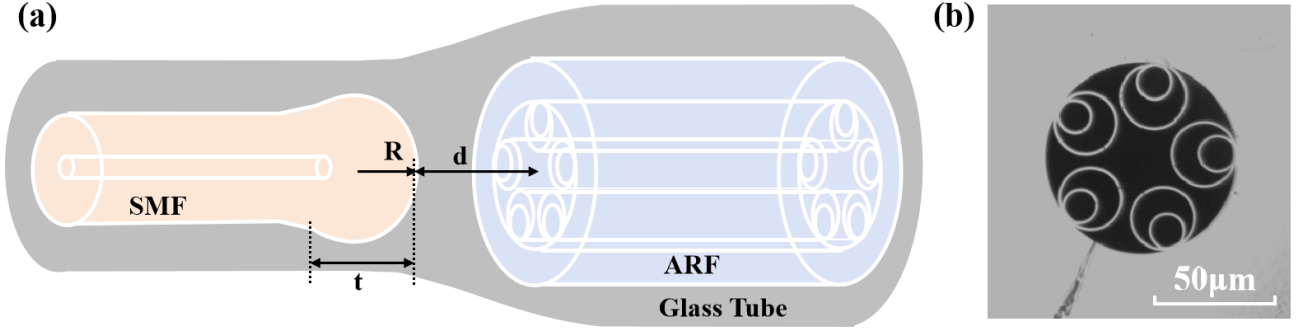

Fig. 1. (a) Schematic diagram of the SMF-ARF coupling structure. (b) Cross-section of the ARF.

To systematically quantify coupling loss, an analytical framework based on normalized electromagnetic field vectors[19] is introduced. And the coupling loss is then calculated by the following equation:

$$\alpha = 1 - \frac{|\int \mathbf{E_1} \cdot \mathbf{E_2} dA|^2}{\int |\mathbf{E_1}|^2 dA \int |\mathbf{E_2}|^2 dA} \quad (1)$$

where $\mathbf{E_1}$ and $\mathbf{E_2}$ represent the electric field vectors of the fundamental modes in the incident and the transmit fibers, respectively. $A$ denotes the cross-sectional area of the fiber. The coupling efficiency between SMF and ARF is calculated to be -2.92 dB, which is prohibitively low for practical applications. Thus, a coupling strategy based on lensed SMF is occupied to address the mismatch. Through structural optimization, the beam divergence from the SMF can be matched to the MFD of the receiving ARF. The strategy makes the coupling loss only relate to the mode-shape mismatch and the Fresnel reflection at the glass-air interfaces, which are minor factors to the coupling efficiency. Remarkably, back-reflection at air-glass interfaces induces interference with forward-propagating signals, a critical consideration in transmission systems. Therefore, the curved lens surface and controlled lens-to-ARF separation in our design are utilized to suppress back-reflection further by angularly deflecting Fresnel-reflected light away from the ARF core axis as shown in Fig. 3. The Fresnel reflection equations are defined as:

$$R_s = \left| \frac{n_1 \cos\theta_1 - n_2 \cos\theta_2}{n_1 \cos\theta_1 + n_2 \cos\theta_2} \right|^2 \quad (2)$$

$$R_p = \left| \frac{n_2 \cos\theta_1 - n_1 \cos\theta_2}{n_2 \cos\theta_1 + n_1 \cos\theta_2} \right|^2 \quad (3)$$

$$R = \frac{R_s + R_p}{2} \quad (4)$$

Where $\theta_1$ and $\theta_2$ represent the angle of incidence and the angle of refraction, respectively; $R_s$ and $R_p$ denote the reflectance for s-polarized and p-polarized light; and $R$ is the total reflectance, representing the fraction of incident light reflected. Although increased angles of incidence for some rays lead to higher reflectivity, the combined effects of lens tilt and offset further attenuate the back-reflection. Overall, the structure allows simultaneous optimization of both coupling efficiency and back-reflection via parameter tuning.

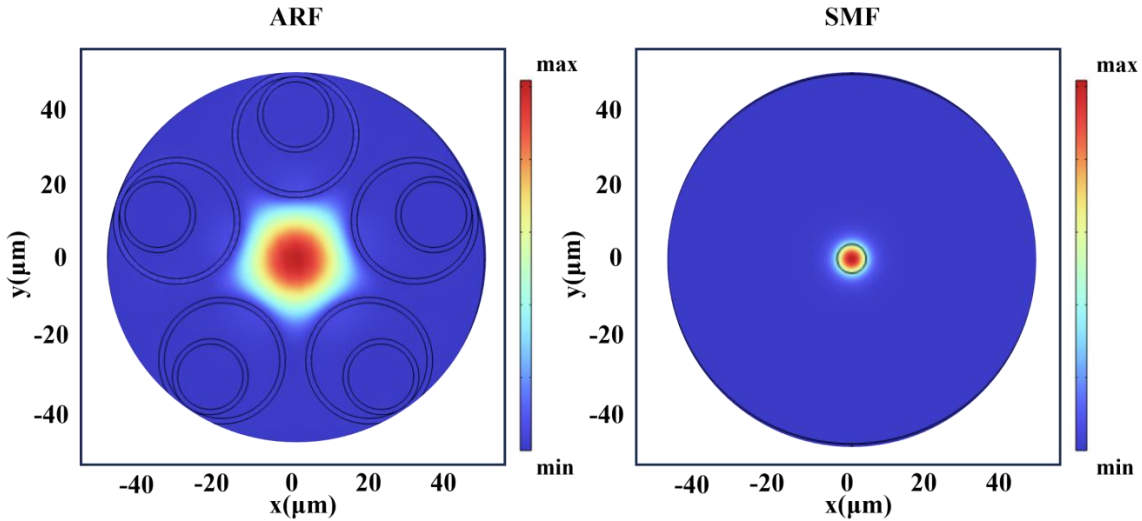

Fig. 2. Mode fields of ARF and SMF.

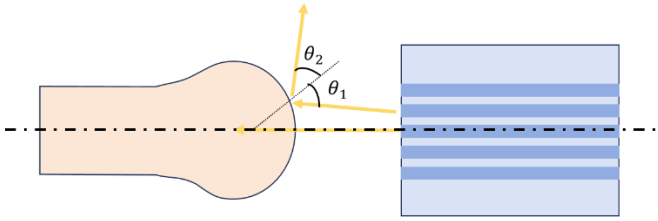

Fig. 3. Off-axis deflection of Fresnel-reflected light from the ARF core.

## 3. Simulation

The investigation focuses on how geometric parameters of the coupling structure and the separation distance between ARF and SMF affect coupling efficiency and back-reflection. As illustrated in Fig. 1(a), mode field matching and back-reflection characteristics depend on three critical parameters: the anterior curvature radius of the lensed SMF (R), the ARF-SMF separation distance (D), and the lens thickness (T). Consequently, since the MFD is significantly smaller than the lens dimensions, the effective curvature radius is defined as that of the lens front surface during fabrication. These parameters affect the coupling efficiency and the intensity of back-reflection to varying degrees, and their combined effects exhibit complex interactions, thus complicating the structural optimization process.

Utilizing the model depicted in Fig. 1(a), simulations are performed using the Zemax Physical Optics Propagation tool to establish the relationship between parameters and performance, as shown in Fig. 4. The ternary contour map illustrates the relationship between coupling efficiency and back reflection in the fiber lens parameter space. Fig 4(a) illustrates parameter-dependent variations in coupling efficiency, revealing distinct peaks resulting from the combined effects of the three variables. Fiber coupling simulations yield a minimum insertion loss of -0.53 dB through optimization at a separation of 342.1 μm, a radius of curvature of 58.7 μm, and a lens thickness of 244.2 μm. The corresponding back-reflection response to these parameters, obtained via non-sequential ray tracing in Zemax, is presented in Fig. 4(b), showing a back-reflection level of -12.32 dB under these optimal conditions. Although this back-reflection level is relatively high, analysis of both Fig. 4(a) and (b) reveals operating points that further optimize the balance between coupling efficiency and back-reflection, even if this entails a slight reduction in coupling efficiency.

Consequently, the impact of offset and tilt on coupling efficiency and back-reflection is simulated to further optimize the balance between these two metrics. The analysis also provides valuable insights for mitigating fabrication errors in experimental device preparation. Leveraging the rotational symmetry of the fiber, the effects of y-axis offset and x-axis tilt are modeled. As shown in Fig. 5, offset significantly degrades coupling efficiency despite marginally reducing back-reflection. Conversely, tilt primarily increases back-reflection with a negligible impact on coupling efficiency within a small angular range. Therefore, under the condition of optimal coupling efficiency, tilting the lens by 6° increases the insertion loss by merely 0.0017 dB while reducing back-reflection to -16.38 dB, thereby achieving an optimized balance between these two metrics.

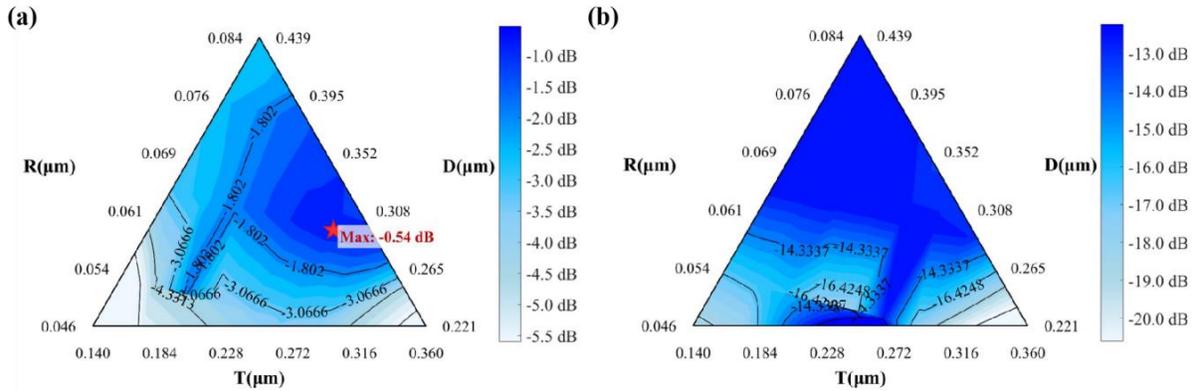

Fig. 4. Ternary contour map of (a) coupling efficiency and (b) back-reflection in fiber lens parameter space.

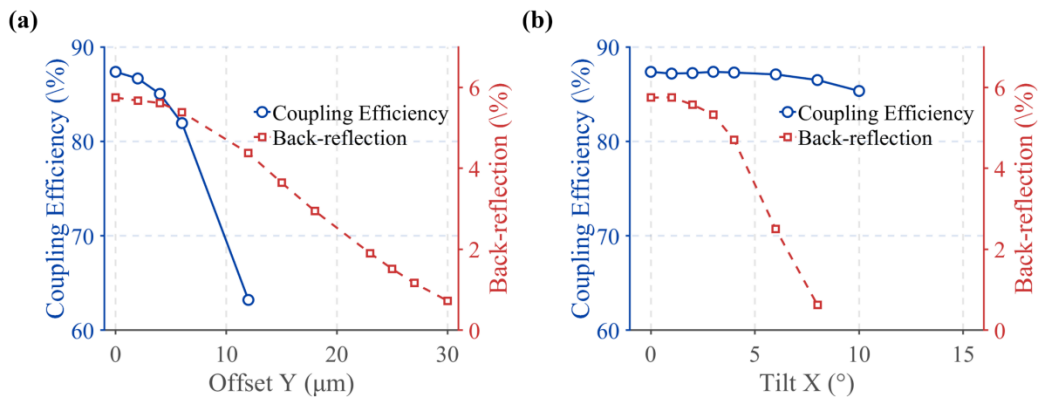

Fig. 5. Relationships between coupling efficiency, back-reflection and (a) offset; (b) tilt.

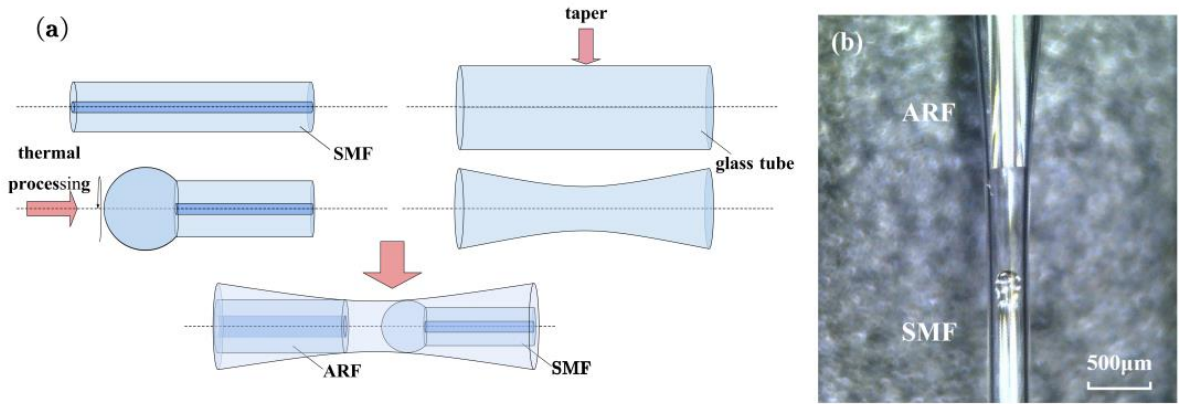

Fig. 6. (a) Fabrication process of the coupling structure. (b) Photograph of the fabricated SMF-ARF coupling structure.

## 4. Experimental results and discussion

### 4.1 Fabrication

The fabrication process comprises three critical steps as illustrated in Fig. 6(a). Firstly, in order to fabricate a hemispherical lens structure with precisely controlled curvature radius, the end of the SMF is thermally processed using the $CO_2$ laser fiber splicing and processing platform LZM-100 through systematic optimization of laser power and motorized pulling speed. This controlled fabrication process enables repeatable production of lens surfaces with sub-micrometer geometric accuracy. Secondly, the alignment process employs a custom-developed assembly fixture, which is realized by transforming a glass tube into a precision alignment sleeve to match the lens dimensions through a controlled tapering process by using a Vytran GPX-3400 glass processing system. Ultimately, the ARF and lensed SMF are inserted into opposite ends of the tapered tube, with the descending inner diameter ensuring stable positioning, as shown in Fig. 6(b). Fine adjustments are made while monitoring IL in real time until optimal coupling efficiency is achieved. Once aligned, ultraviolet adhesive is applied to permanently fix the components, ensuring long-term stability without compromising optical performance.

### 4.2 Measurement

Following the fabrication process of the SMF-ARF coupling structure, we proceeded to characterize its optical performance using the experimental setup illustrated in Fig. 7, which was specifically designed to measure both the IL and back-reflection intensity of the fabricated device. A temperature-stabilized 1550 nm distributed feedback laser serves as the light source, connected through the first port of a high-isolation optical circulator to ensure measuring IL and back-reflection intensity at the same time. The coupling structure is connected to the second output port of the circulator, with its output end linked to a power meter (PM1) for optical power measurement to determine the coupling efficiency. The measured IL is 1.2 dB, corresponding to 75.86% coupling efficiency, which is approximately consistent with the simulation results. Simultaneously, the reflected light from the coupling structure passes through the second port and exited via the third port of the circulator. By connecting the PM to the third port of the circulator, the reflection intensity is measured as -36.22dB, demonstrating superior back-reflection performance. The discrepancies between experimental and simulated results likely originate from fabrication and alignment errors. The fabricated end face deviates from a standard curved surface, causing alterations in the light reflection path. Additionally, insufficient precision in controlling the tilting and misalignment of the ball lens during back-reflection measurements contributes to these differences. Experimental results demonstrate that the proposed configuration simultaneously achieves high coupling efficiency and exceptionally low back-reflection, proving the feasibility of the proposed SMF-ARF structure.

To characterize the wavelength-dependent coupling efficiency of the coupling structure, we replaced PM2 in Fig. 7 with an optical spectrum analyzer and substituted the conventional fiber laser with a supercontinuum source covering 1500 nm to 1600 nm. The output power from ARFs as the coupling input power and the output power from the SMFs through the ball were measured and normalized to the power loss at 1550 nm, as shown in Fig. 8. The results confirm that the 1500-1600 nm wavelength range falls within the operational transmission window of the ARFs, with the coupling efficiency demonstrating a peak-to-peak variation of merely 0.72 dB, which attests to outstanding spectral stability.

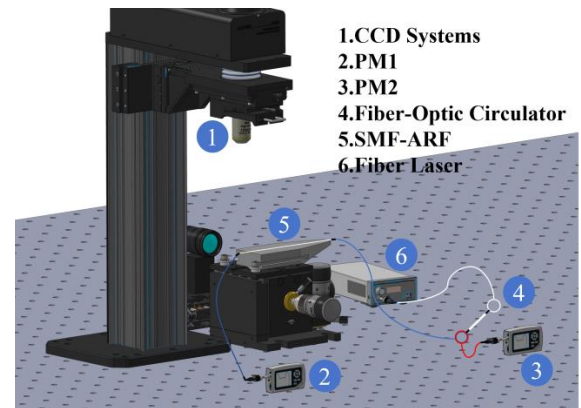

Fig. 7 The coupling efficiency and back-reflection intensity measurement of the coupling structure system experimental setup. (SMF: single mode fiber; ARF: anti-resonant fiber; PM: power meter)

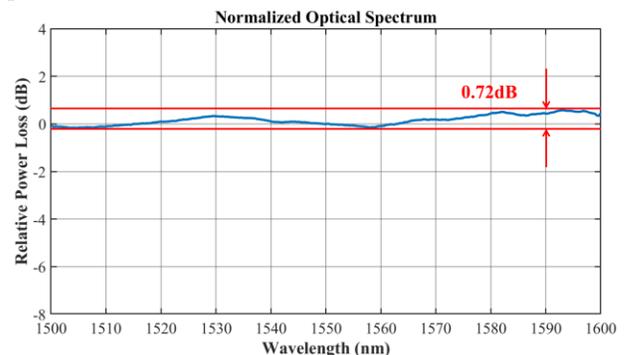

Fig. 8. Normalized optical spectrum of the SMF-ARF coupling. structure.

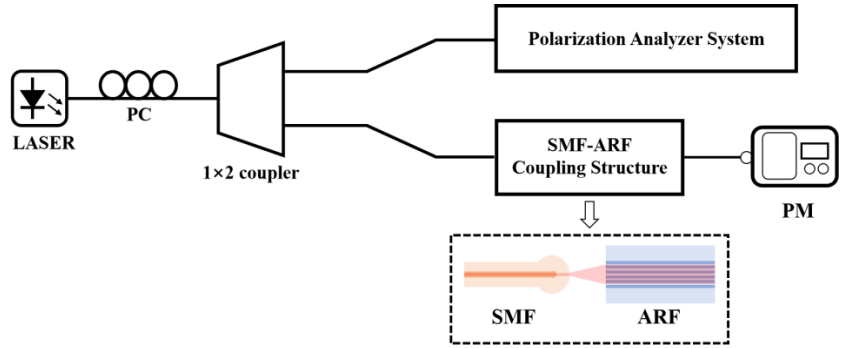

Fig. 9. Polarization measurement system experimental setup. (SMF: single mode fiber; ARF: anti-resonant fiber; PM: power meter; PC: polarization controller)

Meanwhile, the variation in coupling efficiency of the coupling structure under different polarization states was also measured. The polarization measurement system for characterizing SMF-ARF coupling performance, depicted in Fig. 9, is composed of a temperature-stabilized 1550 nm DFB laser followed by a polarization controller (PC) to generate adjustable input polarization states. The light is then split via a 1×2 fiber coupler, with one arm routed to a polarization detection system for real-time monitoring of the polarization state on the Poincaré sphere, while the other arm couples into the SMF-ARF structure where a power meter (PM) measures the transmitted power to determine polarization-dependent coupling efficiency. The system is designed to minimize measurement uncertainty during loss characterization, with its modular configuration and real-time monitoring capability contributing to accurate and reliable characterization of the SMF-ARF coupling performance.

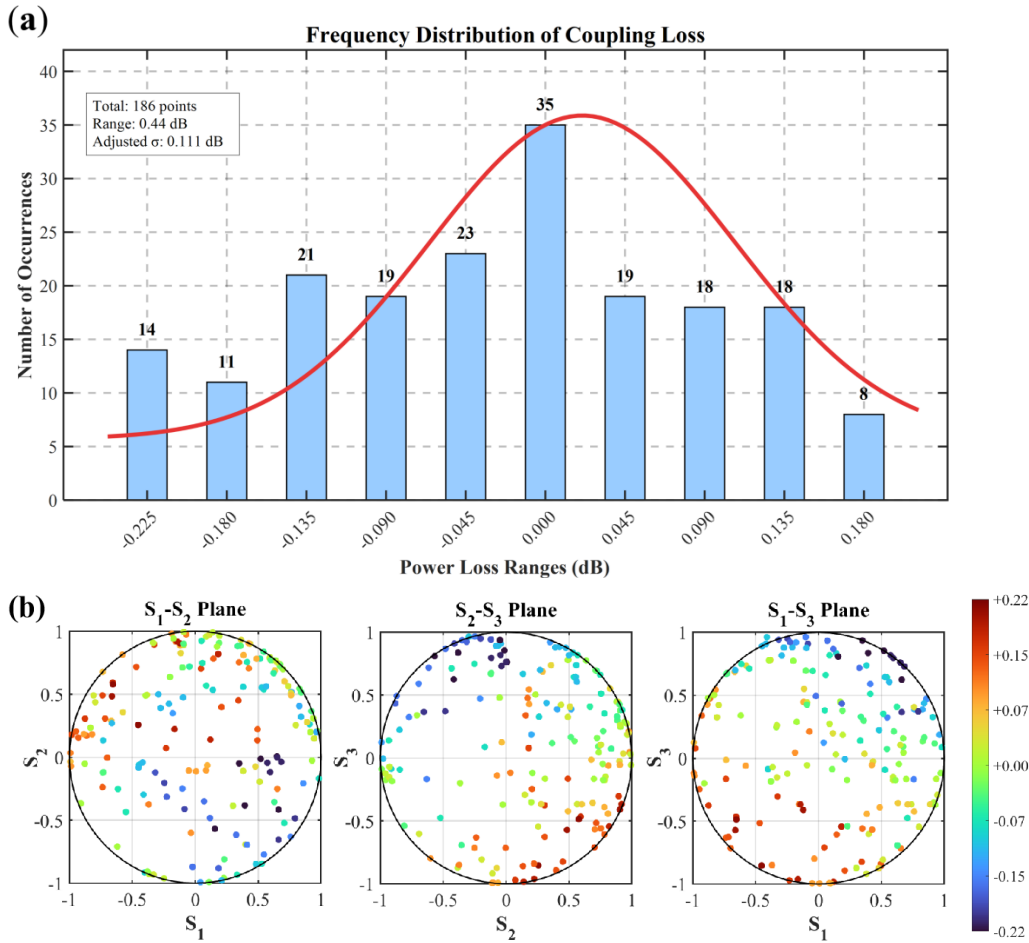

Fig. 10. (a) Frequency distribution of coupling loss. (b) 2D mapping results of the Poincaré sphere.

The statistical distribution of coupling efficiency loss across different Poincaré sphere coordinates is presented in Fig. 10(a), revealing a correlation between loss values and occurrence frequency that resembles a normal distribution with a standard deviation of 0.111 dB. The maximum frequency consistently appears at the central loss value, with a gradual decrease towards both extremes. As demonstrated in Fig. 10(b) through Poincaré sphere mapping, the coupler possesses minimal PDL across all polarization states, confirming the robustness of our design against input polarization variations. The measured coupling efficiency maintains fluctuations within ±3% regardless of polarization angle, demonstrating negligible polarization sensitivity. Analysis of the 2D mapping results of the Poincaré sphere reveals that peak loss fluctuations predominantly occur at the polar regions in both the S2-S3 and S1-S3 planes, suggesting a subtle correlation between loss and polarization states. This phenomenon may originate from either imperfect circumferential symmetry of the lens structure or localized stress birefringence introduced during the thermal lens fabrication process[20]. In practical fiber-optic communication systems where PDL below 1 dB is generally considered acceptable, such level of fluctuation is entirely negligible[21].

## 5. Conclusion

In this work, we proposed and fabricated an easy-to-manufacture lensed SMF-based coupling structure for low-loss and low-reflection SMF-ARF interconnections. The proposed architecture achieves exceptional performance metrics, demonstrating an insertion loss of just 1.2 dB while simultaneously suppressing back-reflection to -36.22 dB which is a significant improvement over conventional coupling schemes matching the simulated predictions. With simplified fabrication, a compact form factor, and excellent spectral stability showing less than 0.72 dB variation across the 1500 nm to 1600 nm wavelength range, the solution ensures reliable broadband operation. Additionally, the polarization-insensitive performance is further proved by measuring the PDL below 0.4 dB. By enabling seamless integration of ARFs into existing optical networks, this advancement addresses the escalating bandwidth and efficiency demands of modern data centers and high-speed communication systems, providing a critical step toward next-generation ultra-high-capacity fiber-optic infrastructure.

## Acknowledgements

This work was supported by the National Key R&D Program of China (2021YFB2800902); National Natural Science Foundation of China (62225110); Major Program (JD) of Hubei Province (2023BAA013).

## References

1. Pei, J. et al. Towards artificial general intelligence with hybrid Tianjic chip architecture. *Nature* **572**, 106–111 (2019).
2. Jadeja, Y. & Modi, K. Cloud computing—concepts, architecture and challenges. In: 2012 International Conference on Computing, Electronics and Electrical Technologies (ICCEET) 877–880 https://doi.org/10.1109/ICCEET.2012.6203873 (2012).
3. Xiong, J., Hsiang, E.-L., He, Z., Zhan, T. & Wu, S.-T. Augmented reality and virtual reality displays: emerging technologies and future perspectives. Light Sci. Appl. 10, 216 (2021).
4. M. Patterson et al., Carbon Emissions and Large Neural Network Training, arXiv:2104.10350 (2021).
5. Dayarathna, M., Wen, Y. & Fan, R. Data center energy consumption modeling: a survey. *IEEE Commun. Surv. Tutor.* **18**, 732–794 (2016)
6. Cheng, Q., Bahadori, M., Glick, M., Rumley, S. & Bergman, K. Recent advances in optical technologies for data centers: a review. OPTICA 5, 1354–1370 (2018).
7. Index, Global Cloud . "Cisco Global Cloud Index: Forecast and Methodology, 2016 – 2021 White Paper.". https://www.cisco.com/c/dam/global/en_au/assets/cisco-live/ites2013mel/assets/presentations/Cloud_Index_White_Paper.pdf
8. Agrawal, Govind P. Fiber-optic communication systems. John Wiley & Sons, 2012.
9. M.Michieletto, J. K. Lyngsø, C. Jakobsen, J. Lægsgaard, O. Bang, and T. T. Alkeskjold, "Hollow-core fibers for high power pulse delivery," Opt. Express 24(7), 7103–7119 (2016).
10. Poletti, F., Wheeler, N., Petrovich, M. et al. Towards high-capacity fibre-optic communications at the speed of light in vacuum. Nature Photon 7, 279 – 284 (2013). https://doi.org/10.1038/nphoton.2013.45
11. Pryamikov A D, Gladyshev A V, Kosolapov A F, et al. Hollow-core optical fibers: current state and development prospects[J]. Physics-Uspekhi, 2024, 67(2): 129-156.
12. Dmytro Suslov, Eric Numkam Fokoua, Daniel Dousek, Ailing Zhong, Stanislav Zvánovec, Thomas D. Bradley, Francesco Poletti, David J. Richardson, Matěj Komanec, and Radan Slavík, "Low loss and broadband low back-reflection interconnection between a hollow-core and standard single-mode fiber," Opt. Express 30, 37006-37014 (2022)
13. Y. Min et al., "Fusion Splicing of Silica Hollow Core Anti-Resonant Fibers With Polarization Maintaining Fibers," in Journal of Lightwave Technology, vol. 39, no. 10, pp. 3251-3259, 15 May15, 2021, doi: 10.1109/JLT.2021.3058888.
14. N. Zhang, Z. Wang, Y. Chen and X. Xi, "Low-loss coupling by injecting tapered solid-core fibers into the hollow-core of anti-resonant fibers," 2017 16th International Conference on Optical Communications and Networks (ICOCN), Wuzhen, China, 2017, pp. 1-3, doi: 10.1109/ICOCN.2017.8121343.
15. B. Shi, C. Zhang, E. R. Numkam Fokoua, F. Poletti, D. J. Richardson and R. Slavík, "Towards spliced SMF to hollow core fiber connection with low loss and low back-reflection," 2023 Conference on Lasers and Electro-Optics (CLEO), San Jose, CA, USA, 2023, pp. 1-2.
16. Y. Jung, J. Hayes, Y. Sasaki, K. Aikawa, S. U. Alam, and D. J. Richardson, "All-fiber optical interconnection for dissimilar multicore fibers with low insertion loss," in Proc. of OFC 2017, paper W3H.2.
17. D. Suslov et al., "Low loss and high-performance interconnection between standard single-mode fiber and antiresonant hollow core fiber," Sci. Rep., vol. 11, no. 1, Apr. 2021, Art. no. 8799.
18. Yablon, A. D. Optical Fiber Fusion Splicing; Springer: Germany, 2005.
19. Aghaie, K. Z.; Digonnet, M. J.; Fan, S. Optimization of the splice loss between photonic-bandgap fibers and conventional singlemode fibers. Opt.Lett. 2010, 35, 1938-1940.
20. Y. Chen, A. Yi, L. Su, F. Klocke, and G. Pongs, "Numerical simulation and experiment study of residual stresses in compression molding of precision glass optical components", J. Manuf. Sci. Eng. 130, 051012 (2008).
21. M. Shtaif and A. Andrusier, "Polarization dependent loss and polarization mode dispersion in coherent polarization multiplexed transmission," 2012 Asia Communications and Photonics Conference (ACP), Guangzhou, China, 2012, pp. 1-